\documentclass[a4paper,fleqn,usenatbib]{mnras}

\usepackage{amsmath}	

\usepackage[T1]{fontenc}
\usepackage{ae,aecompl}

\usepackage{graphicx}	

\usepackage{amssymb}	

\title[The photo-chemical evolution of the Sculptor dSph]{Lighting up stars in chemical evolution models: the CMD of Sculptor}
\author[F. Vincenzo et al.]{F. Vincenzo$^{1,2}$\thanks{E-mail:
vincenzo@oats.inaf.it}, F. Matteucci$^{1,2,3}$, T.~J.~L. de Boer$^{4}$, M. Cignoni$^{5}$ and M. Tosi$^{6}$
\\
$^{1}$Dipartimento di Fisica, Sezione di Astronomia, Universit\`a di Trieste, via G.B. Tiepolo 11, 34100, Trieste, Italy\\
$^{2}$INAF, Osservatorio Astronomico di Trieste, via G.B. Tiepolo 11, 34100, Trieste, Italy\\
$^{3}$INFN, Sezione di Trieste, Via Valerio 2, 34100, Trieste, Italy\\
$^{4}$Institute of Astronomy, University of Cambridge, Madingley Road, Cambridge CB3 0HA, UK\\
$^{5}$Space Telescope Science Institute, 3700 San Martin Drive, Baltimore, MD, 21218, USA\\
$^{6}$INAF, Osservatorio Astronomico di Bologna, Via Ranzani 1, I-40127, Bologna, Italy  }

\begin{document}

\date{Accepted 2016 May 11. Received 2016 May 11; in original form 2016 February 2}

\pagerange{\pageref{firstpage}--\pageref{lastpage}} \pubyear{2016}

\maketitle

\label{firstpage}


\begin{abstract}
We present a novel approach to draw the synthetic color-magnitude diagram of galaxies, which can 
provide -- in principle -- a deeper insight in the interpretation and understanding of current observations.  
In particular, we `light up' the stars of chemical evolution models, according to their initial mass, metallicity and age, to eventually 
understand how the assumed underlying galaxy formation and evolution scenario affects the final configuration of the synthetic CMD.  
In this way, we obtain a new set of observational constraints for chemical evolution models beyond the usual photospheric 
chemical abundances. 
The strength of our method resides in the very fine grid 
of metallicities and ages of the assumed database of stellar isochrones. 
In this work, we apply our photo-chemical model to reproduce the observed CMD of the Sculptor dSph 
and find that we can reproduce the main features of the observed CMD. 
The main discrepancies are found at fainter magnitudes in the main sequence turn-off and sub-giant branch, 
where the observed CMD extends towards bluer colors 
than the synthetic one; we suggest that this is a signature of metal-poor stellar populations in the data, 
which cannot be captured by our assumed one-zone chemical evolution model.

\end{abstract}



\section{Introduction}

In this work, we present a novel approach to obtain a synthetic color-magnitude diagram (CMD) of galaxies, starting from predictions of 
chemical evolution models. 
Our new \textit{photo-chemical model} `lights up' 
the stars of chemical evolution models, according to their initial mass, metallicity and age; 
in this way, we can obtain a new set of observational constraints for chemical evolution models
beyond the usual photospheric chemical abundances. 
The method presented in this work can provide -- in principle -- a deeper insight in the interpretation 
of current observations, since we can understand how our hypothesis about galaxy formation and evolution can affect the final configuration 
of the CMD. 

\par By solving a set of physically-motivated differential equations, which take into account the main physical processes taking place and influencing the evolution of the galaxy interstellar medium (ISM), numerical codes of chemical evolution are able to provide the galaxy star formation 
history (SFH) and age-metallicity relation; the evolution of the galaxy stellar and gas mass, and the run of the ISM chemical abundances with time. 
Building up a photo-chemical code consists then in coupling the output of chemical evolution models with a database of stellar isochrones, 
currently available and computed with very high accuracy.


\begin{figure}
\includegraphics[width=9.3cm]{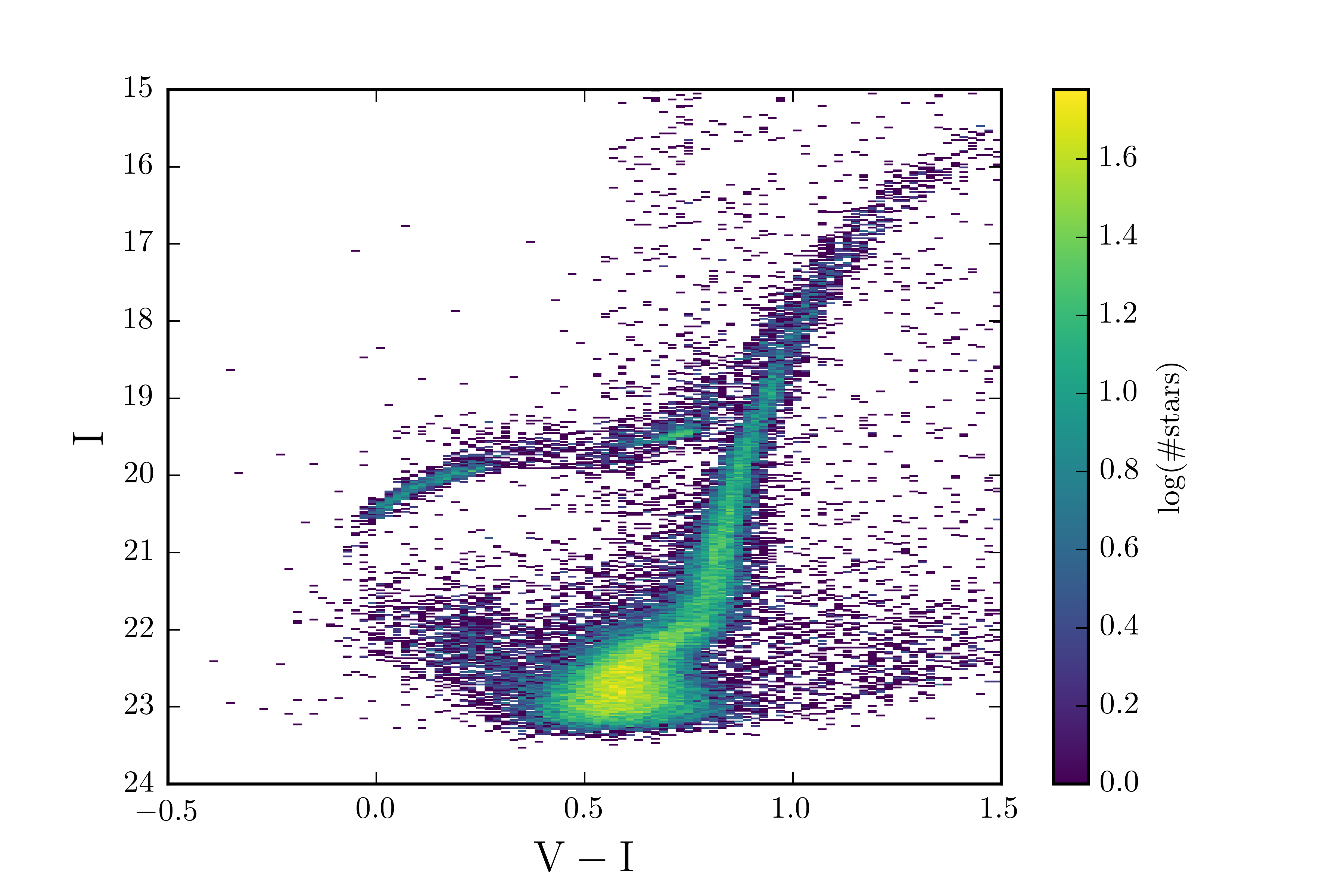}
\caption{  In this Figure, we show the observed CMD of the Sculptor dSph \citep{deboer2011}. The dataset is shown as 2-D histogram, 
with the bin size in both the $x$- and $y$-dimensions being $0.02\,\mathrm{dex}$; 
the color coding in the figure corresponds to the number of 
stars within each grid element. We consider in our analysis only clean isolated 
stellar detections and the Sculptor member stars with an elliptical radius $r_{\mathrm{ell}}\leq0.6\,\mathrm{deg}$.  
 }
     \label{fig:CMD_observed}
\end{figure} 


\par Most of the previous works in the literature recover the SFH of galaxies from the observed CMD 
by adopting sophisticated fitting techniques (e.g. \citealt{harris2001,dolphin2002,aparicio2004,tolstoy2009,cignoni2010,monelli2010,hidalgo2011}); 
in particular, they search for the suitable linear combination of simple stellar populations (SSPs) 
with different age and metallicity, which provides the best agreement with the observed photometric properties of the galaxy composite stellar population. 
As a byproduct, this `classical' procedure can also predict an average galaxy age-metallicity relation. 
Nevertheless, no underlying approximate physical model is assumed in these works for the galaxy formation and evolution. 

\par In this paper, the first of a series of future works, we focus on reproducing the CMD of the Sculptor 
dwarf spheroidal galaxy (dSph). 
In particular, we investigate whether the best chemical evolution model for Sculptor -- reproducing the galaxy stellar metallicity distribution function (MDF) 
 -- 
is able to predict a synthetic CMD which agrees with the observed one.

 \par This work is organized as follows: in Section \ref{sec:data} we summarize the main properties of the Sculptor dSph and describe the observed set of data 
 used in this work for the comparison with our models; in Section \ref{sec:model} we present the main characteristics of our 
 photo-chemical model and the methods we employ to fairly compare the synthetic with the observed CMD; 
 in Section \ref{sec:results} we show our results, and in Section \ref{sec:conclusion} we draw some conclusions. 

\section{The observed dataset} \label{sec:data}
The Sculptor dwarf galaxy was discovered by \citet{shapley1938}. 
Although it might appear simple at first glance, from the study of the kinematical, chemical and 
spatial distribution of its 
stellar populations, \citet{tolstoy2004} were able to disentangle in this galaxy an inner, kinematically `cold', metal-rich stellar 
population from an outer `hot' metal-poor one, later on confirmed by \citet{battaglia2008} and \citet{walker2011}. 
Other studies based on photometric datasets also were able to identify (or confirm) the existence of stellar populations distinct in metallicity 
\citep{majewski1999}, age and spatial distribution \citep{deboer2011,deboer2012}. 

\par \citet{mcconnachie2012} reported for Sculptor an average V-band surface brightness 
$\mu_{V}=23.5\pm0.5\,\text{mag arcsec}^{-2}$, an half-light radius $r_{\mathrm{h}}=283\pm45\,\mathrm{pc}$, and an absolute V-band magnitude $M_{\mathrm{V}}=-11.1\pm0.5\,\mathrm{mag}$. 
We make use of the distance modulus $\mu=19.62\pm0.04\,\mathrm{mag}$ 
derived by \citet{martinez2015}. 

\par The observed CMD is taken from \citet[see Fig. \ref{fig:CMD_observed}]{deboer2011}, 
which were able to resolve stars down to the oldest main sequence turn-off (MSTO) of the 
Sculptor dSph, by taking advantage of the deep wide-field photometry of CTIO/MOSAIC. 
In order to avoid a non-negligible contamination of foreground MW disc field stars, 
we consider only stars along the line-of-sight to the Sculptor dSph with elliptical radius $r_{\mathrm{ell}}\leq0.6\,\mathrm{deg}$ and 
corresponding to clean isolated detections; 
the percentage of stars with these characteristics in the \citet{deboer2011} catalog is about $92$ per cent of the entire sample.  
We find that, for $r_\text{ell} > 0.6\,\text{deg}$, the noisy pattern introduced by foreground stars becomes 
larger than the ``signal'' of the Sculptor CMD one wants to recover (see also figure 5 in \citealt{deboer2011}). 

\par The observed stellar MDF is taken from \citet{romano2013}, 
which combined the \citet{kirby2009,kirby2010} spectroscopic sample (with Sculptor stars belonging to the inner $0.2\,\text{deg}$ 
of elliptical radius) with the dataset provided by the Dwarf galaxies Abundances and Radial velocities 
Team (DART, covering a much larger radial extent and making use of the 
calcium triplet equivalent width to infer the 
Fe abundances; see \citealt{battaglia2008,starkenburg2010}), in order to have an MDF which were 
representative of the global Sculptor stellar populations.  
By looking at \citet[their figure A1]{romano2013}, the low-metallicity portion of their MDF 
is almost solely determined by the DART sample, with the \citet{kirby2009,kirby2010} MDF mainly contributing 
towards larger [Fe/H] abundances.


\begin{figure}
\includegraphics[width=9.3cm]{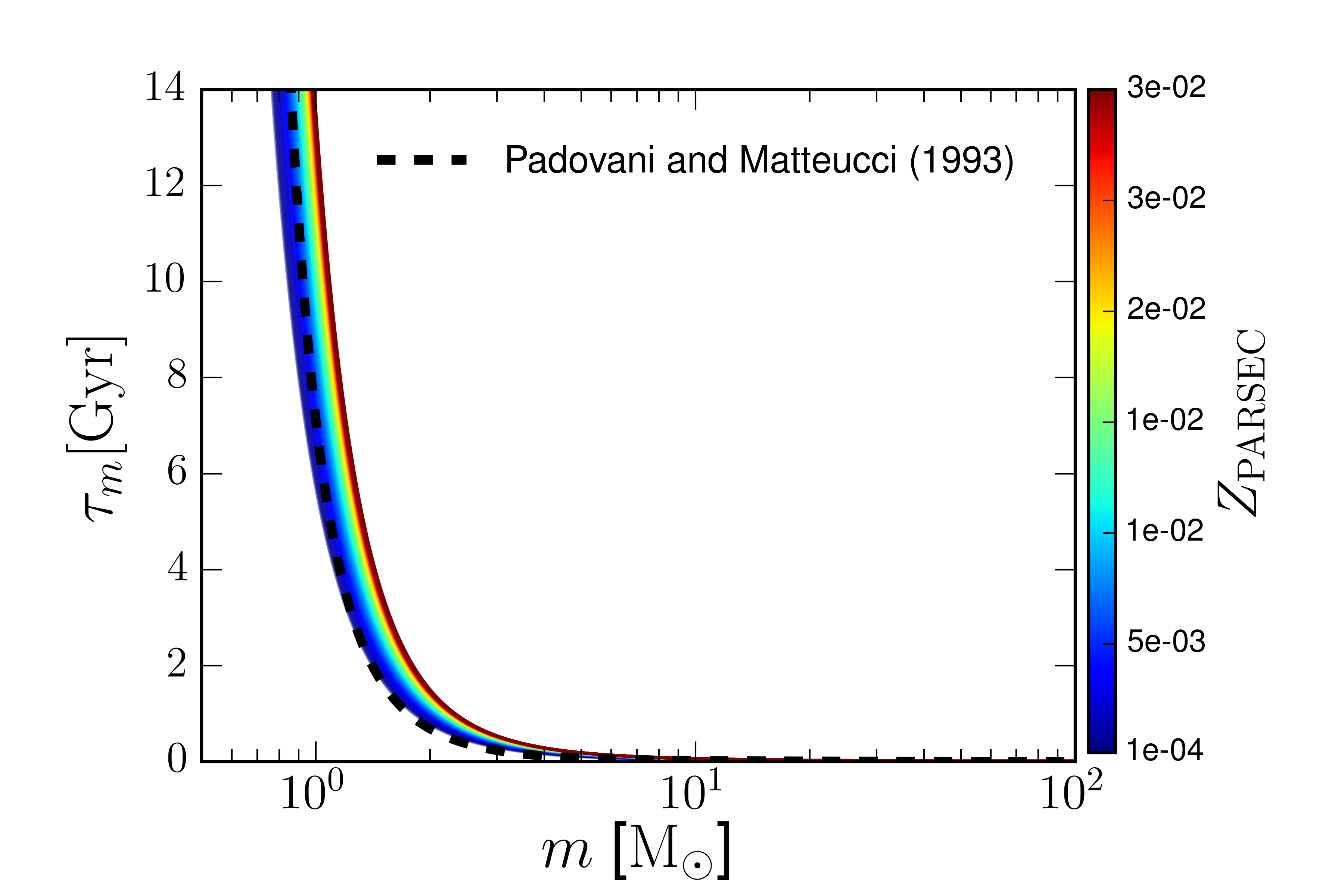}
\caption{ In this figure, we show how the stellar lifetimes we have derived from the PARSEC stellar evolutionary tracks vary as functions of the 
stellar mass and metallicity. The dashed black curve corresponds to the stellar lifetimes of \citet{padovani1993}. 
}
     \label{fig:lifetimes}
\end{figure} 


\section{Model, assumptions and methods}  \label{sec:model}

\subsection{Database of stellar isochrones and stellar lifetimes} \label{sec:lifetimes}


\begin{figure}
\includegraphics[width=9.3cm]{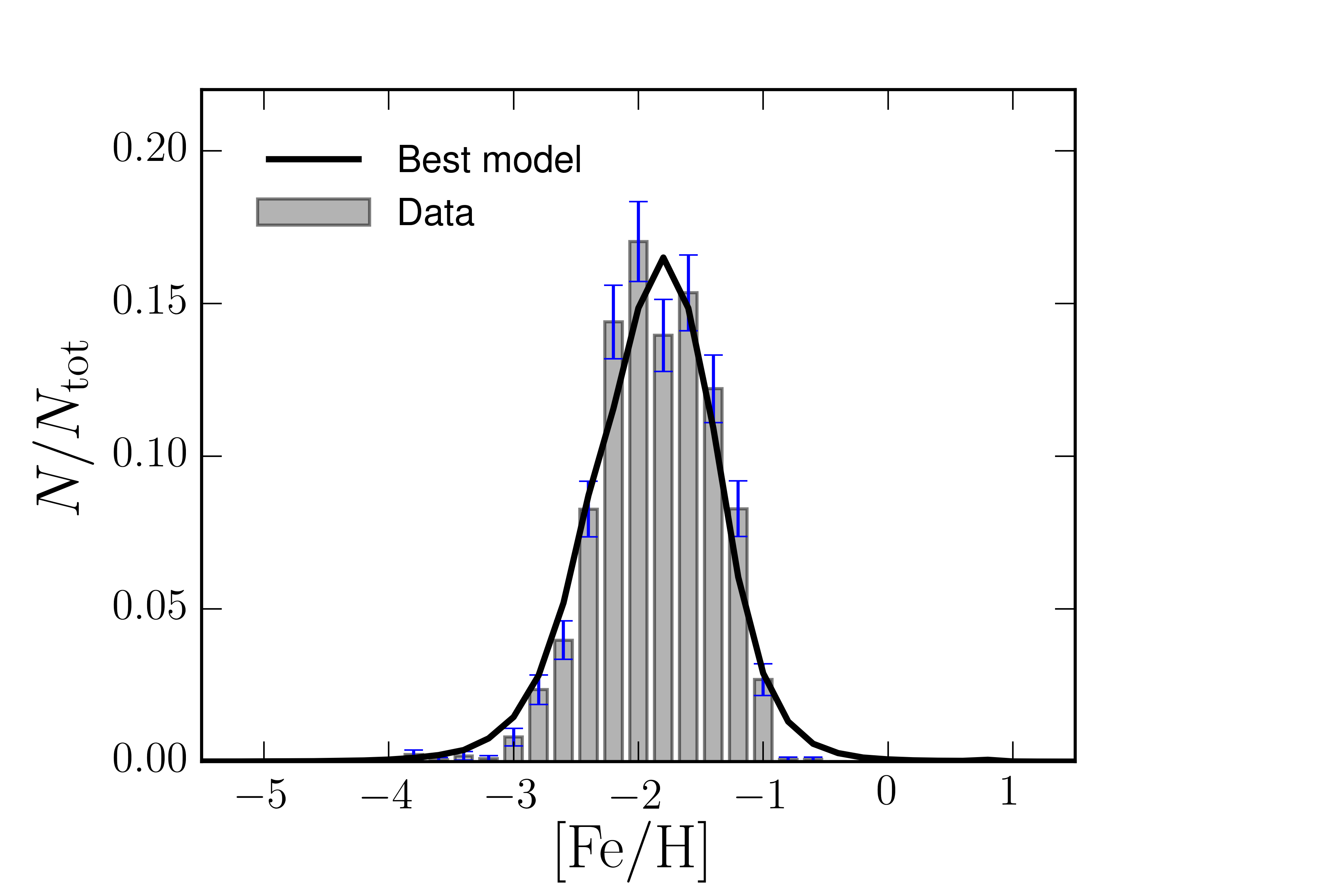}
\caption{ In this figure, we compare the observed Sculptor stellar MDF \citep[grey histogram with blue error bars]{romano2013} with the predictions 
of our best Sculptor chemical evolution model, having star formation efficiency $\nu=0.04\,\mathrm{Gyr}^{-1}$, wind efficiency 
$\lambda_{\mathrm{wind}}=3.0\,\mathrm{Gyr}^{-1}$, infall mass $M_{\mathrm{inf}}=2.31\times10^{8}\,M_{\mathrm{\sun}}$, and infall 
time-scale $\tau_{\mathrm{inf}}=0.3\,\mathrm{Gyr}$. The predicted MDF is convolved with a Gaussian function with $\sigma=0.2$. 
}
     \label{fig:mdf}
\end{figure} 


We make use of the PARSEC stellar isochrones \citep{bressan2012,tang2014,chen2015}, as computed for the following grid of 
stellar ages and metallicities, by assuming a Reimers mass loss with efficiency $\eta=0.2$.  
\begin{enumerate} 
\item The step in metallicity in our isochrone database is $\Delta{Z}=1.0\times10^{-4}$, from a minimum metallicity 
$Z_{\mathrm{min}}=1.0\times10^{-4}$ to a maximum metallicity $Z_{\mathrm{max}}=3.0\times10^{-2}$.  
\item The step in age between two adjacent isochrones is $\Delta\log({t/\mathrm{yr}})=0.01$, 
from a minimum age $\log({t_{\mathrm{min}}/\mathrm{yr}})=6.5$ 
to a maximum age $\log({t_{\mathrm{max}}/\mathrm{yr}})=10.12$. 
\end{enumerate} 
For self-consistency, we assume in our model the same stellar lifetimes as the ones which can be derived from the PARSEC database; 
in particular, we fit the stellar lifetimes with the following function: 
\begin{equation} \tau_{m}(Z) = \mathrm{A}(Z) \times \exp{\left[\mathrm{B}(Z)\,m^{-\mathrm{C}(Z)} \right]},   \end{equation}
where $\mathrm{A}(Z)$, $\mathrm{B}(Z)$ and $\mathrm{C}(Z)$ are the fitting parameters, provided with the corresponding 
$1$-$\sigma$ errors in the supplementary material, as functions of the metallicity $Z$. 
In Fig. \ref{fig:lifetimes} we compare our derived stellar lifetimes with the ones of \citet{padovani1993}, which do not depend on metallicity 
and are extensively used in chemical evolution models.

 
\subsection{Modelling the chemical evolution of Sculptor}

The numerical code of chemical evolution is the same as the one 
adopted in \citet{vincenzo2014,vincenzo2015} -- where we address the reader for details -- for the study of the 
classical and ultra-faint dSph galaxies. 
We make use of an updated version, by assuming the stellar yield compilation of \citet[their model 15]{romano2010} 
and the stellar lifetimes derived from the PARSEC isochrones. 
\par We assume the galaxy to assemble by accreting pristine gas from an external reservoir, until an infall mass -- given by 
$M_{\mathrm{inf}}$ -- is accumulated at 
$t_{\mathrm{G}}=14\,\mathrm{Gyr}$. The infall rate is assumed to follow a decaying exponential law, with typical time-scale $\tau_{\mathrm{inf}}$. 
We assume for the star formation rate the Schmidt-Kennicutt law, namely $\mathrm{SFR}(t)=\nu M_{\mathrm{gas}}(t)$, 
where $\nu$ is the so-called 
star formation efficiency (SFE) and $M_{\mathrm{gas}}$ is the galaxy gas mass. 
The run of the intensity of the SFR with time is crucially regulated by the various physical processes acting on  $M_{\mathrm{gas}}$, 
namely inflows and outflows of gas, returned matter from dying stars and supernovae,  
astration due to the star formation activity itself. 

\par A fundamental role in the evolution of dSphs is played by the galactic outflows, 
which are predicted to occur very soon in these galaxies 
because of their shallow potential well; 
the intensity of the outflow rate is assumed to be directly proportional to the SFR. 
On the one hand, if the efficiency for the gas removal is high (typically $\lambda_{\mathrm{wind}}\approx10\,\mathrm{Gyr}^{-1}$), 
then the galaxy gas mass suddenly 
decreases and hence the SFR rapidly drops to zero; on the other hand, if the galactic wind has a relatively lower efficiency 
(typically $\lambda_{\mathrm{wind}}\approx1\,\mathrm{Gyr}^{-1}$), then the decrease in the SFR is smoother and it drops to zero on longer 
typical time-scales. 


\begin{figure}
\includegraphics[width=9.3cm]{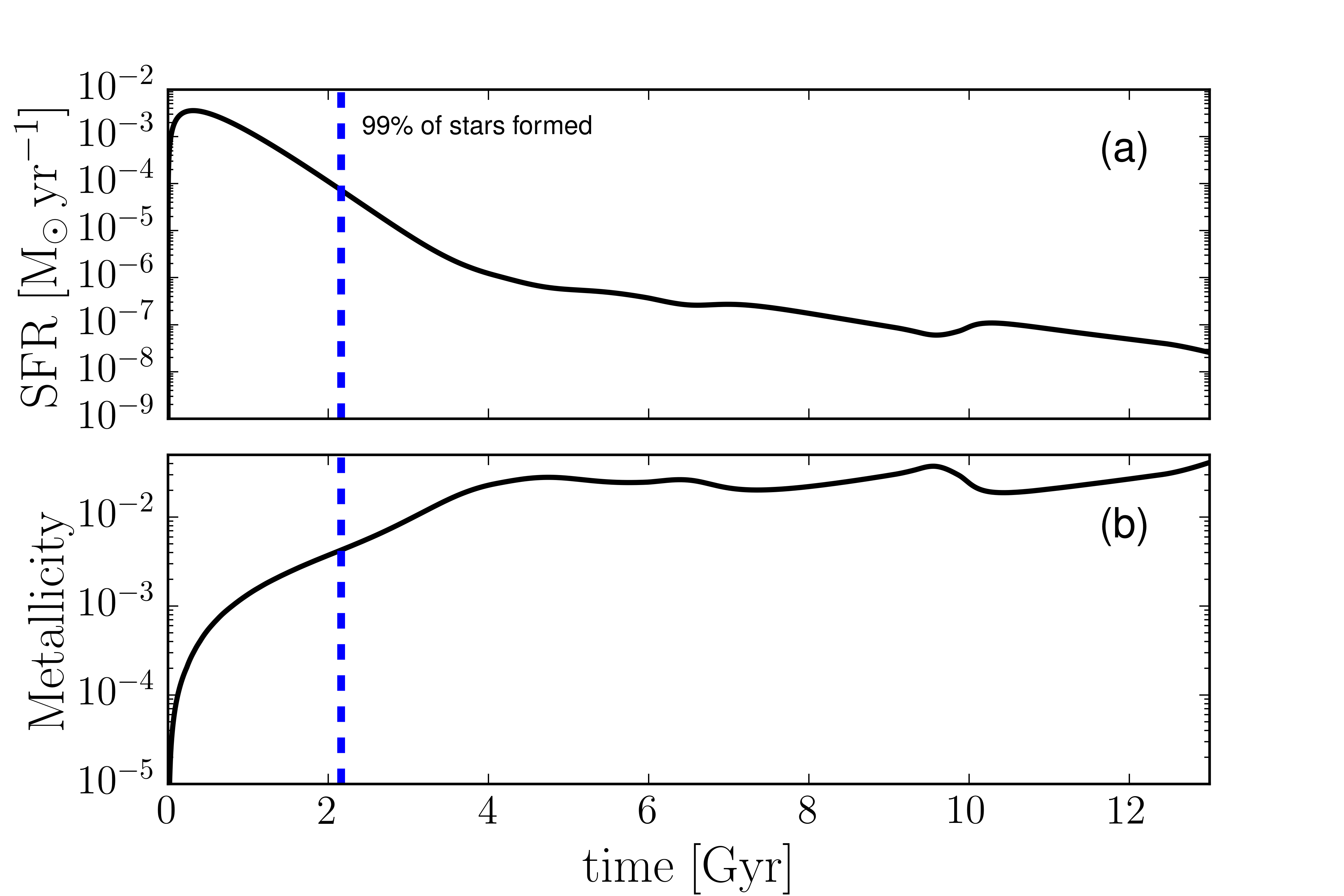}
\caption{ In this figure, we show the predicted SFH (top panel) and age-metallicity relation (bottom panel) as predicted by our 
best chemical evolution model for Sculptor. Our best model for Sculptor predicts that $\sim99$ per cent of the stars 
observable at the present time are formed within the first $2.16\,\mathrm{Gyr}$ of the galaxy evolution; this time 
corresponds to the vertical dashed blue line in the figures. Furthermore, 
the number of stars with initial metallicity $Z<1.0\times10^{-4}$ is roughly $\sim5.72$ per cent of the total number of stars 
 alive at the present time. 
 }
     \label{fig:SFH}
\end{figure} 


\subsubsection{The assumed best model}


We have explored the parameter space, by running a large number of chemical evolution models. 
The best parameters for Sculptor are found by minimizing the $\chi^{2}$ figure of merit, with the best model being the one 
reproducing  the shape of the observed stellar MDF, which represents the most reliable 
observational constraint to any slight variation of the free parameters. 
We vary the SFE in the range 
$\nu=0.03$-$0.2\,\mathrm{Gyr}^{-1}$, the wind efficiency in the range  $\lambda_{\mathrm{wind}}=2.0$-$10\,\mathrm{Gyr}^{-1}$, 
and the infall time-scale in the range $\tau_{\mathrm{inf}}=0.1$-$0.5\,\mathrm{Gyr}$. 
In order to allow our photo-chemical model to be more flexible when comparing its predictions with data, we 
make the further approximation of assuming the galactic wind to be always active over the whole galaxy lifetime; 
since all the galaxy physical properties in our chemical evolution model are normalized with respect to the infall mass,  
this assumption causes the predicted stellar MDF not to be influenced by the variation of the infall mass. In 
this simplified formulation of the galactic wind, all the galaxy physical properties (such as the SFR, stellar mass, gas mass, 
and so on) simply scale with the infall mass and our results can be easily readjusted for a different assumption of the cutoff in the 
elliptical radius of Sculptor stars (here we consider only stars with $r_{\text{ell}} \le 0.6\,\text{deg}$). 
We find that the best chemical evolution model is characterized by the following parameters: 
\begin{enumerate}
\item star formation efficiency $\nu=0.04\,\mathrm{Gyr}^{-1}$; 
\item wind efficiency $\lambda_{\mathrm{wind}}=3.0\,\mathrm{Gyr}^{-1}$; 
\item infall time-scale $\tau_{\mathrm{inf}}=0.3\,\mathrm{Gyr}$. 
\end{enumerate} 
We assume at the beginning a reference infall mass $M_{\text{inf,ref}} = 1.0\times10^{8}\,\mathrm{M}_{\sun}$, 
as in \citet{vincenzo2014}, for which we predict a present-day total stellar mass $M_{\star,\text{ref}}=8.27\times10^{5}\,\mathrm{M}_{\sun}$.  
The infall mass of the best model is then obtained by rescaling our results for the reference model 
so as to have the same number of stars 
in the synthetic and observed CMD. We find for our best model an infall mass 
$M_{\text{inf,best}} = 2.31\times10^{8}\,\mathrm{M}_{\sun}$, giving rise to a present-day total stellar 
$M_{\star,\text{best}}=1.91\times10^{6}\,\mathrm{M}_{\sun}$, larger than the value 
$M_{\star}=( 1.2 \pm 0.6 ) \times 10^{6}\,\text{M}_{\sun}$ estimated by \citet{deboer2012}, but of the same order of magnitude.

\par According to the fitting formula of \citet{faucher2011}, which 
is assumed in many recent works to mimic a cosmologically motivated infall in galaxy formation and evolution models, 
dwarf galaxies with $M_\text{halo} =10^{8}\,\text{M}_{\sun}$ 
must have accreted almost $63$ per cent of their cumulative infall mass (which turns out to be 
$M_{\text{inf}}\approx1.4\times10^{8}\,\text{M}_{\sun}$) 
in the first $\sim 1.5\,\text{Gyr}$ of their evolution, 
namely from redshift $z=6$ to redshift $z\sim4$, a larger timescale than the one found by our best model 
($\tau_\text{inf} = 0.3\,\text{Gyr}$). Nevertheless, we remark on the fact that the 
\citet{faucher2011} fitting formula is strictly valid only in the redshift range $0 < z < 6$, namely 
only after the reionization epoch; therefore, a comparison with the gas mass assembly history 
of our model is not straightforward. 

\par In Fig. \ref{fig:mdf} the predicted stellar MDF of the best model (convolved with 
a Gaussian function with $\sigma=0.2\,\mathrm{dex}$) is compared with the observed one \citep{romano2013}. 
The assumed very low SFE causes the MDF to be peaked at low $[\mathrm{Fe/H}]$ abundances. The width of the MDF is mostly determined 
by the wind efficiency; 
in particular, the lower the $\lambda_{\mathrm{wind}}$ parameter, the wider is the bulk of the galaxy star formation activity 
and hence also the MDF.

\par In Figure \ref{fig:SFH}a) we show the predicted SFH of our best model, while in Figure \ref{fig:SFH}b) we show the corresponding 
age-metallicity relation. In summary, the length of the bulk of the galaxy star formation activity can be regulated in our model  
by suitably varying the main parameters determining the star formation and outflow intensity and the galaxy gas accretion rate; 
these parameters are the 
SFE, which determines the intensity of the SFR and the rate of thermal energy injection by supernovae,  
the wind efficiency, which determines the slope with which the SFR drops to zero, and the infall time-scale, which crucially 
determines the evolution with time of the galaxy potential well.

\subsection{The photo-chemical model}

\par The stepwise structure of the photo-chemical model is the following. 
\begin{enumerate}
\item We sample the galaxy SFH, as predicted by our best chemical evolution model for Sculptor, to randomly extract an age for the formation of a 
given star.  
\item We sample the assumed initial mass function (IMF) to randomly assign a mass to the star. In this work we assume the \citet{salpeter1955} IMF 
for simplicity.
\item We use the age-metallicity relation of our best Sculptor chemical evolution model 
to find the initial metallicity of the star. 
\item Given the age, mass and metallicity of the star, we check whether the star is alive or not at the present time, by assuming the metallicity 
dependent stellar lifetimes we have derived from the PARSEC stellar evolutionary tracks (see Section \ref{sec:lifetimes}). 
\item If the star can be observed at the present time, we store the photometric properties of the synthetic star, to later draw it in the synthetic CMD. 
\end{enumerate} On the one hand, the strength of our method resides in the very fine grid of the assumed isochrone database; moreover, in our 
approach, we start from the predictions of chemical evolution models, assuming \textit{ab initio} an underlying galaxy formation and evolution scenario, 
which is physically-motivated. 
On the other hand, the main shortcoming of our model is due to the fact that the lowest available metallicity in the PARSEC database is 
$Z_{\mathrm{min}}=1.0\times10^{-4}$. We assume that all the stars with $Z < Z_{\mathrm{min}}$ have the same photometric properties as the stars with $Z=Z_{\mathrm{min}}$. This fact can introduce a systematic error. 
By looking at Figs. \ref{fig:SFH}a) and b), according to our best model, the galaxy spends its first 
$122\,\mathrm{Myr}$ at metallicity $Z<1.0\times10^{-4}$; 
the number of stars 
with initial metallicity $Z<1.0\times10^{-4}$ is roughly $\sim5.72$ per cent of the total number of stars 
 alive at the present time. 

\subsubsection{Methods}


\begin{figure*}
\includegraphics[width=\textwidth]{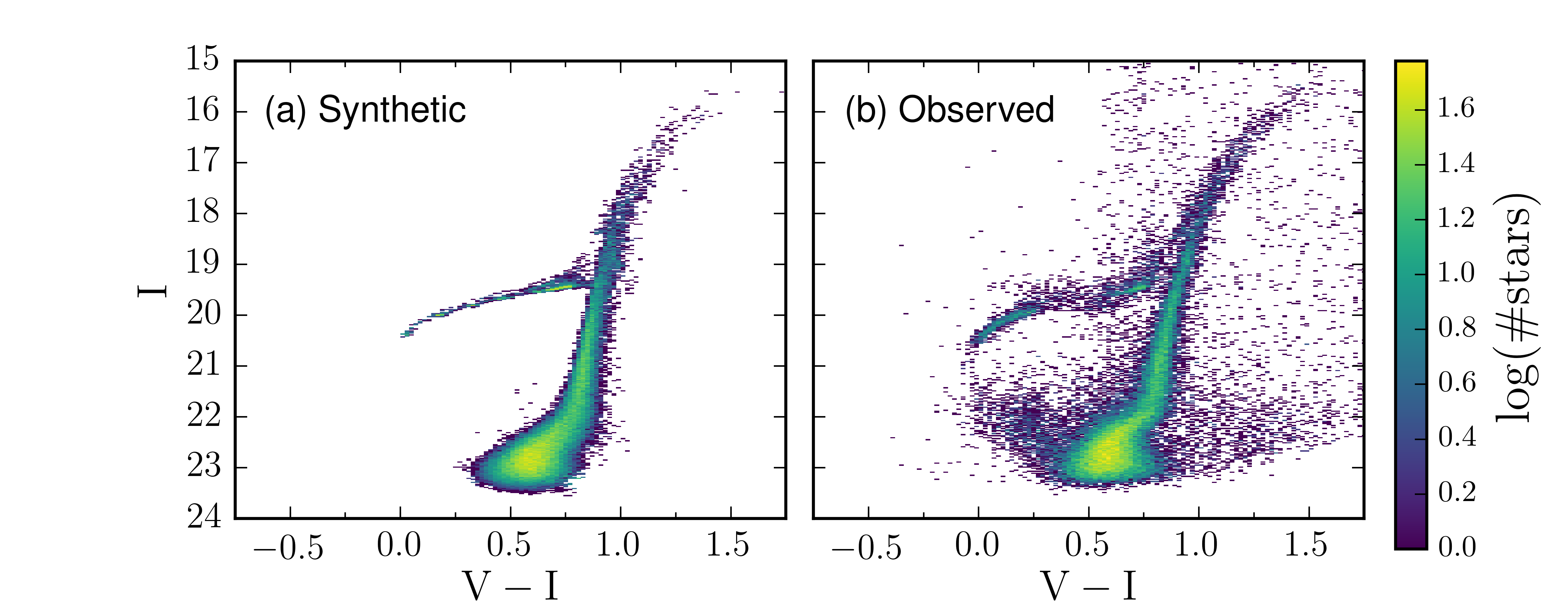}
\caption{  In the left panel of this figure, we show the prediction of our photo-chemical model for the CMD of the Sculptor dSph, whereas on the 
right panel we show for comparison the observed CMD. The synthetic and the observed CMDs are shown as 2-D histograms, with the bin size  
in both the $x$- and $y$-dimensions being $0.02\,\mathrm{dex}$ and 
the color-coding representing the number of stars, on a logarithmic scale, residing 
within each grid element. 
 }
     \label{fig:CMD_result}
\end{figure*} 


\par To get a fair comparison with the observed CMD, we convolve the synthetic CMD with the distribution of the observed photometric errors, 
by assuming that the latter are Gaussian. In particular, we first divide the observed CMD in an uniform grid and, 
for each grid element $ij$, we compute the average V- and I-band observed photometric errors, 
$\overline{\sigma}_{ij}(\mathrm{V,I})$, 
which we adopt as the standard deviations of the photometric noise in the $ij$-th grid element. 
Hence, for any given $k$-th synthetic star residing in the $ij$-th grid element, we add the following noise  
to its predicted V- and I-band magnitudes: 
$ \overline{\sigma}_{k}(\mathrm{V,I}) = r_{k}\times\overline{\sigma}_{ij}(\mathrm{V,I}) $, 
where $r_{k}$ is a random number, 
drawn according to the standard normal distribution. In this way, the model `spreads out' 
according to the errors in the data and we can fairly compare the synthetic with the observed CMD. 
\par The synthetic CMD corrected for the photometric noise is then convolved with the results of the artificial star test performed by \citet{deboer2011}.   
In particular, by following a standard procedure, \citet{deboer2011} inserted in the observed images a large catalog of artificial stars; 
after reducing and 
analyzing the altered images, they could compute the fraction of artificial stars recovered in the data as a function of their input magnitude and color. 
We use the results of their test to compute the recovered fraction in each grid element, so as  
to throw out from the synthetic CMD the remaining lost fraction.  By means of this type of analysis, one can apply to the synthetic CMD the same 
completeness profile as is present in the data. 
After correcting the synthetic CMD for the incompleteness, the number of synthetic stars strongly reduces, becoming 
$N_{\mathrm{tot,syn}}\approx40636$.

\section{Results}  \label{sec:results}

\begin{figure}
\includegraphics[width=9.3cm]{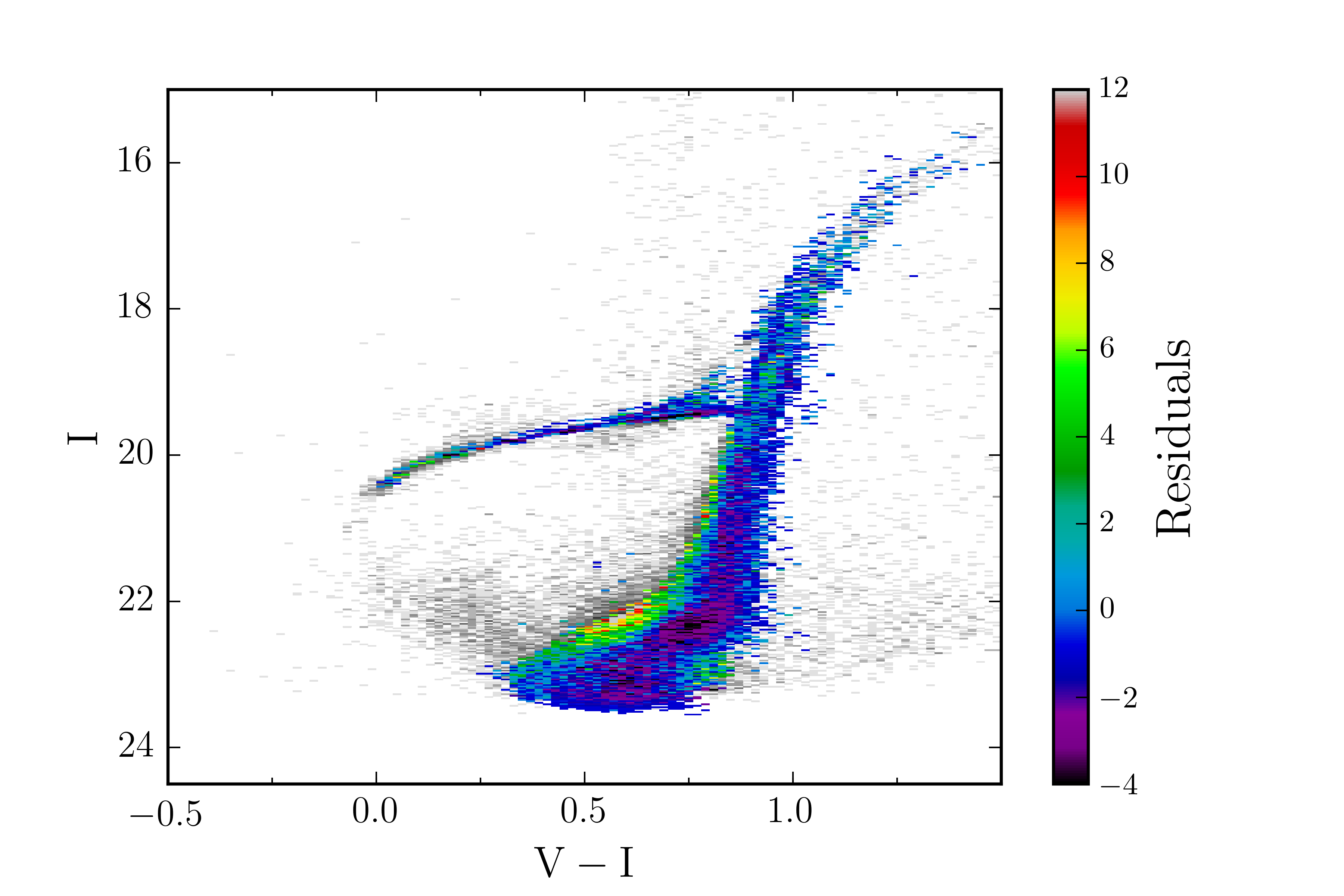}
\caption{  In this Figure, we show the residuals (see eq. \ref{eq:res} ) for the comparison between the observed and the synthetic CMD. The greyscale 
density plot represents the regions of the observed Sculptor CMD where no synthetic stars are predicted. 
We find a good agreement 
for the RGB, HB and AGB stars, whereas the MSTO and the SGB of the observed CMD extend towards bluer colors than the synthetic CMD. 
We cannot reproduce the population of blue straggler stars in the observed CMD, since we do not include them in our model. 
 }
     \label{fig:residuals}
\end{figure}

In this Section, we present the results of our photo-chemical model for the CMD of the Sculptor dSph. 
The main result of our work in shown in Fig. \ref{fig:CMD_result}, where the synthetic CMD of Sculptor (left diagram) is compared 
with the observed one (right diagram). 
In order to better understand where the discrepancies between the observed and the synthetic CMD reside, in Fig. 
\ref{fig:residuals} we plot the residuals, which correspond to the color-coding in the figure. In particular, 
to better visualize the differences, we define the residual in the $ij$-th grid element as: 
\begin{equation} \label{eq:res} R_{ij}= \frac{n_{ij,\mathrm{obs}}-n_{ij,\mathrm{syn}}}{\sqrt{n_{ij,\mathrm{syn}}}}, \end{equation} with 
$n_{ij,\mathrm{obs}}$ and $n_{ij,\mathrm{syn}}$ being the number of stars in the observed and synthetic CMD, respectively. The 
regions of the observed Sculptor CMD without any synthetic star are shown in Fig. \ref{fig:residuals} as a greyscale density plot.  


\par On the one hand, by a visual inspection of Figs. \ref{fig:CMD_result} and \ref{fig:residuals}, 
we can obtain a quite good agreement for the red giant branch (RGB), 
the horizontal branch (HB) and 
the asymptotic giant branch (AGB) of the observed CMD. 
On the other hand, at fainter magnitudes, particularly in the sub-giant branch (SGB) and at the MSTO, 
the observed CMD extends towards slightly bluer colors than the synthetic one. 
Moreover, our model cannot reproduce the observed population of blue straggler stars which extend the Main Sequence towards blue colors 
and could be -- in principle -- reproduced by including the effect of merging binary stellar systems. We do not include blue straggler stars 
in our photo-chemical model. 


\par We remark on the fact that it is not obvious that a model reproducing the chemical evolution of Sculptor can also capture the main features of the 
observed galaxy CMD. 
In fact, the final configuration of the synthetic CMD turns out to be highly affected by the variation of the free parameters of chemical evolution models. 
In particular, by increasing the SFE, the stellar metallicities accordingly increase at any fixed galactic time, causing the entire synthetic CMD to shift 
towards redder colors. The IMF acts in a similar way as the SFE, with the additional effect of filling up the various stellar evolutionary phases in the CMD 
with different relative fractions. 
Finally, the wind parameter and the infall time-scale crucially affect the spread of the predicted CMD, since they determine the length of the bulk of the galaxy 
star formation activity.

\par
\begin{figure}
\includegraphics[width=9.3cm]{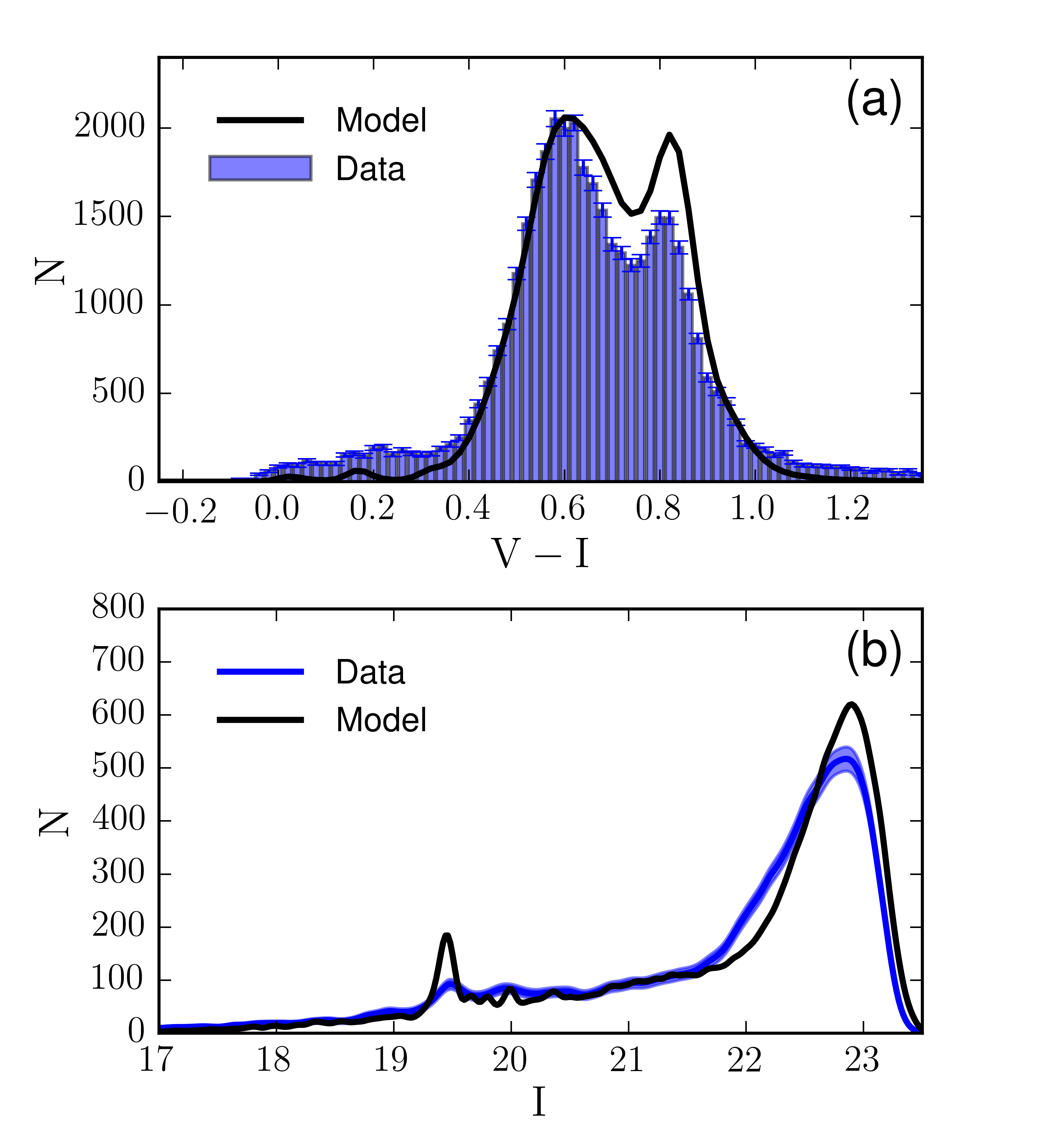}
\caption{  In the top panel of this figure, we compare the predicted stellar $(V-I)$-color distribution (black solid line) with the observed one 
(blue histogram with errorbars). 
In the bottom panel, we compare the predicted stellar luminosity function in the $I$-band (back solid line) with the observed one (blue line, with the 
shaded blue area representing the $1$-$\sigma$ errors). MSTO stars are predicted to reside in the synthetic CMD at $m_{I}\ga\,23.0\,\mathrm{mag}$, 
whereas SGB stars at $22.0 \la m_{I}\la\,23.0\,\mathrm{mag}$. The peak in the $I$-band stellar luminosity function 
at $m_{I}\approx\,19.5\,\mathrm{mag}$ is caused by HB stars. 
 }
     \label{fig:VI_histo}
\end{figure}

 In Fig. \ref{fig:VI_histo}a), we compare the predicted stellar $(V-I)$-color distribution (black solid line) with the observed one (blue histogram); 
this quantity turns out to be particularly sensitive to metallicity variations among the galaxy stellar populations. 
In Fig. \ref{fig:VI_histo}b), the predicted stellar luminosity function in the $I$-band (black solid line) is compared with the observed one 
(blue line with error bars, which are shown as a shaded blue area); the trend of this second quantity is more affected by the galaxy SFH and 
stellar lifetimes. 
An age indicator for the galaxy is given by the fraction of 
stars on the HB relative to the one on the RGB; we predict $N_{\mathrm{HB}}/N_{\mathrm{RGB}}\approx0.17$.


\par At a first glance, our photo-chemical model is able to predict and qualitatively reproduce the main features of the 
observed $(V-I)$-color distribution; in particular, the first peak in Fig. \ref{fig:VI_histo}a) (the one at bluer colors) is determined by the MSTO and SGB stars, 
whereas the second peak (the one at redder colors) is the signature left by the ascending RGB and HB stars.

\par Concerning the blue portion of the color distribution, from a visual inspection of Fig. \ref{fig:VI_histo}a), 
we cannot reproduce the observed population of blue straggler stars, which -- as aforementioned -- are not included in our photo-chemical model. 
Furthermore, a remarkable discrepancy resides in the decaying trend of the blue wing of the predicted color distribution, 
which contains a lower number of stars than the data, and in the predicted `saddle', which 
turns out to be higher than the observed one. 
This can be also appreciated by looking at the residual plot in Fig \ref{fig:residuals}, where the observed CMD clearly contains a larger number of 
MSTO and SGB 
stars with blue colors than the synthetic CMD. This discrepancy seems likely the signature of metal-poor stellar populations in the Sculptor dSph, 
which our one-zone chemical evolution model has not been able to capture. Nevertheless, in principle, it could also indicate a predicted 
age-metallicity relation which is steeper than what seems to be required by observations; in fact, 
one would obtain a similar discrepancy if also the synthetic metal-rich stars are older than the observed ones. 

\par Interestingly, by looking at Fig. \ref{fig:mdf}, the observed stellar MDF suggests the presence of two distinct peaks, corresponding to two 
separated main stellar populations in the galaxy. 
The latter feature cannot be resolved by our best chemical evolution model, which indeed predicts the stellar MDF 
to have a single peak, lying between the two of the observed distribution. We remark on the fact that this feature 
is peculiar to the \citet{romano2013} MDF, which combines the DART sample -- determining the low metallicity portion of the 
observed MDF -- with the one by 
\citet{kirby2009,kirby2010}, concentrated towards slightly higher metallicity. 
The regions where the observed CMD contains a larger number of blue (metal-poor) stars than the synthetic one 
likely correspond to the low-metallicity peak in the observed MDF, which is also the most pronounced one.  
Accordingly, the higher `saddle' in the predicted color distribution (see the top panel in Fig. \ref{fig:VI_histo}) 
confirms that the model predicts galaxy stellar populations which are intermediate 
between the observed metal-poor and metal-rich ones.

\par The observed stellar MDF, as derived by \citet{romano2013}, 
includes stars with $r_{\text{ell}} \ge 0.6\,\text{deg}$, namely beyond the 
radial cutoff we assume for the observed CMD.  As one moves towards the outer 
zones of the galaxy, the Sculptor stellar populations are observed to become increasingly
 old and metal-poor, determining 
 the observed stellar MDF at low metallicity. 
 Nevertheless, the relative number of stars 
 also increasingly diminishes when moving outwards, with almost $50$ per cent of the \citet{romano2013} 
sample being contributed by stars with 
$r_{\text{ell}} \le 0.2\,\text{dex}$.  
Furthermore, the shape of the observed stellar MDF in the inner radial bins 
does not significantly vary at low metallicity with respect to the MDF in the outer bins (see figure 14 of 
\citealt{deboer2012}). Therefore, if we had applied a radial cutoff at $r_{\text{ell}} = 0.6\,\text{deg}$ 
also to the observed MDF, our analysis would not have been significantly altered.

\par A viable solution to reduce the discrepancy between the observed and the synthetic CMD would be, for example, 
to disentangle the two different stellar populations in the observed MDF, by assuming two underlying separated distributions, peaked at slightly 
lower and higher $[\mathrm{Fe/H}]$ abundances than the predicted MDF of the best model. 
In this way, the metal-poor stellar population could be reproduced by assuming a lower SFE than the metal-rich one. 
After superimposing the two stellar populations in the synthetic CMD with appropriate weights, one would extend the synthetic CMD towards slightly 
bluer color, hence obtaining a better agreement with the observed CMD.  
Finally, we cannot exclude that the lack of binary stars in our photo-chemical model might also contribute to 
the discrepancy between the observed and the synthetic CMD, since their inclusion would cause a broadening of the MS and therefore a 
redistribution of the colors of the synthetic stars in Fig. \ref{fig:VI_histo}a).

\par In Fig. \ref{fig:VI_histo}b), the observed 
stellar $I$-band luminosity function is compared with the predicted one. For the RGB and AGB stars, 
there is a good agreement between the model and data. Furthermore, the model predicts a peak at 
$m_{I}\approx\,19.5\,\text{mag}$, which represents the effect of HB stars; the presence of this peak is not visible 
in the observed $I$-band stellar luminosity function because of the large foreground contamination in the redder part of the observed CMD, 
both at fainter and at brighter $I$-band magnitudes than the ones of the observed HB. 
If we had considered only Sculptor stars in the innermost regions of the galaxy 
(e.g. with $r_\text{ell}\le 0.2 \, \text{deg}$), we would have obtained a well agreement between model and data also for the HB stars. 
At fainter magnitudes, the model in Fig. \ref{fig:VI_histo}b) 
contains a larger number of MSTO and a lower number of SGB stars than the data; 
this discrepancy might be partly due to the inherent uncertainty in the assumed completeness profile, which is particularly important 
for these evolutionary stages, as well as it could also indicate the need for an IMF with a lower number of low-mass stars than the 
\citet{salpeter1955} IMF assumed in this work.

\section{Conclusions and discussion}  \label{sec:conclusion}

In this work, we have presented a new approach to draw the synthetic CMD of galaxies. 
In particular, we have developed \textit{ab-initio} a new photo-chemical model, which we have applied to reproduce the observed CMD of Sculptor 
dSph. Our numerical code starts from the predictions of chemical evolution models about the galaxy 
SFH and age-metallicity relation. Then, by assuming  the PARSEC stellar evolutionary tracks, 
we can `light up' the stars with different age, mass and metallicity of chemical evolution models, in order to draw 
a synthetic CMD. 
We have defined the best chemical evolution model for Sculptor as the one reproducing the observed galaxy stellar MDF. 

\par Several improvements could be done in our photo-chemical model, by 
 considering for example an underlying cosmological framework, whose primary effect would be to influence the evolution of the galaxy 
gas mass assembly with time. Interestingly, a very 
first attempt to draw the CMD of a dSph galaxy within a full cosmological framework by making use of a semi-analytical model for the 
galaxy formation and evolution is represented by the work of \citet{salvadori2008}, 
which adopted the freely available IAC-STAR code \citep{aparicio2004}, however, they did not provide a detailed discussion of their 
findings about the predicted galaxy CMD. A further improvement in our photo-chemical model would be to include the 
effects of unresolved binary stars and foreground contamination in the synthetic CMD.  
 
\par The strength of our approach resides in the statistical sampling of the galaxy predicted SFH and assumed IMF, 
as well as in the assumption of a very fine grid of stellar isochrones, both in metallicity and in age. 
The main shortcoming is related to the 
PARSEC stellar isochrones, which are computed only for $Z\ge10^{-4}$.

\par The SFH can be regulated in our models for dSphs by suitably varying the main parameters triggering the onset of the 
galactic wind and determining its intensity. In fact, once the galactic wind has started, the SFR rapidly drops to zero, 
since most of the infall mass has been accumulated fast within short typical time-scales. 
Our best model for Sculptor predicts that $\sim99$ per cent of the stars 
observable at the present time are formed within the first $2.16\,\mathrm{Gyr}$ of the galaxy evolution. 
We predict at the present time a total stellar mass $M_{\star,\text{best}}=1.91\times10^{6}\,\mathrm{M}_{\sun}$, 
which is of the same order of magnitude as other estimates like $M_{\star}=( 1.2 \pm 0.6 ) \times 10^{6}\,\text{M}_{\sun}$ by \citet{deboer2012}. 
Also the predicted evolution of the SFR as a function of time is in agreement with the findings of \citet{deboer2012}. 

\par Stellar systems or interstellar regions with low gas density, such as low-mass dwarf galaxies or the outer parts of spiral galaxies, 
likely follow a star formation law which deviates from the usually assumed Schmidt-Kennicut law,  
$\mathrm{SFR}(t)=\nu M_{\mathrm{gas}}(t)$; 
for this reason, we have done some numerical experiments by assuming the same  
expression for the star formation rate as in the original work of \citet{kennicutt1998} (see also \citealt{gatto2015} for a detailed study 
in the context of hydrodynamical simulations). 
By assuming the \citet{kennicutt1998} law, we predict 
the SFH to be more concentrated in the earliest epoch of the galaxy evolution and the metallicity Z 
to initially evolve more rapidly than our best-fitting model; then, at later times, Z remains quite constant 
when the \citet{kennicutt1998} law is assumed, while it increases in our best-fitting model. 
Finally, we find that the final total gas and stellar mass are almost the same when the two different expressions  
for the star formation rate are assumed. 

\par We have shown that our photo-chemical model is able to capture the main features of the observed CMD of the Sculptor dSph, 
with the best agreement being obtained for the RGB, HB and AGB stars. 
The discrepancy has been found at fainter luminosity in the MSTO and SGB, where the observed CMD 
extends towards bluer colors than the synthetic one. That may be caused by underlying metal-poor stellar populations which our photo-chemical model 
has not been able to capture as well as to the lack of binary stars in our model, which would also broaden the synthetic CMD at faint 
magnitudes. 
In fact, the predicted stellar MDF is characterized by a single peak, 
whereas the observed one suggests the presence of two peaks, residing at slightly lower and higher $[\mathrm{Fe/H}]$ abundances than the model peak. 
In particular, the more pronounced peak in the observed MDF corresponds to the one at lower $[\mathrm{Fe/H}]$ abundances. 
Therefore, our photo-chemical model contains stellar populations which are intermediate between the metal-poor and the metal-rich ones 
in the observed stellar MDF. 

\par In order to reduce the discrepancy, one could superimpose the results of multiple one-zone chemical evolution models 
and find the linear combination which provides the best agreement with the observed stellar MDF. 
This will be the subject of a future work, in which we will also show the 
effect of varying the free parameters of chemical evolution models on the synthetic CMD. 

\par Although the uncertainties in the assumed completeness profile can be important at the MSTO and SGB, the discrepancy between model and data 
in the $I$-band stellar luminosity function for $m_{I} \ga 21.0\,\mathrm{mag}$ 
might be alleviated by assuming an IMF which contains a lower number of low-mass stars than the \citet{salpeter1955} IMF 
assumed in this work and by including the effect of binary stars.


\section*{Acknowledgements}
FV thanks E. Brocato for insightful discussions during the visit at the Astronomical Observatory of Rome in 2014 December, 
and S. Recchi for his precious suggestions. 
FM and MT acknowledge financial support from PRIN-MIUR~2010-2011 project 
`The Chemical and Dynamical Evolution of the Milky Way and Local Group Galaxies', prot.~2010LY5N2T.  
We thank an anonymous referee for his/her constructive comments. 


\bsp

\label{lastpage}

\end{document}